\documentclass{caosp305}

\usepackage{graphicx}

\usepackage{natbib}
\articleNo{}
\pubyear{2017}
\volume{}
\volnumber{}
\firstpage{1}
\received{November 2017}
\accepted{}

\def\BibTeX{{\rm B\kern-.05em{\sc i\kern-.025em b}\kern-.08em
             T\kern-.1667em\lower.7ex\hbox{E}\kern-.125emX}}

\begin{document}

%
\hauthor{Z. Keszthelyi \textit{ et al.}}

\title{The impact and evolution of magnetic confinement in hot stars}


%
\author{
        Z.\,Keszthelyi \inst{1,} \inst{2,}  
      \and 
        G.A.\,Wade \inst{1}   
      \and 
        V.\,Petit \inst{3}
			\and
			  G.\,Meynet \inst{4}
			\and 
			  C.\,Georgy \inst{4}
       }

%
\institute{
           Department of Physics and Space Science, Royal Military College of Canada, PO Box 17000 Station Forces, Kingston, ON, K7K 0C6, Canada \\       \email{zsolt.keszthelyi@rmc.ca}
         \and 
				 Department of Physics, Engineering Physics and Astronomy, Queen's University, 99 University Avenue, Kingston, ON, K7L 3N6, Canada       
         \and
        Department of Physics and Astronomy, University of Delaware, Newark, DE 19716, USA 
         \and
         Geneva Observatory, University of Geneva, Maillettes 51, 1290 Sauverny, Switzerland
         }

\date{November 10, 2017}

\maketitle

\begin{abstract}
Magnetic confinement of the winds of hot, massive stars has far-reaching consequences on timescales ranging from hours to Myr. Understanding the long-term effects of this interplay has already led to the identification of two new evolutionary pathways to form `heavy' stellar mass black holes and pair-instability supernova even at galactic metallicity. We are performing 1D stellar evolution model calculations that, for the first time, account for the surface effects and the time evolution of fossil magnetic fields. These models will be thoroughly confronted with observations and will potentially lead to a significant revision of the derived parameters of observed magnetic massive stars. 
\keywords{stars -- massive -- magnetic}
\end{abstract}

%
\section{Introduction}
\label{intr}
Massive stars are key objects in the Universe, shaping their local environment via chemical enrichment and energy deposition. Therefore it is crucial to understand the physical processes governing the structure and evolution of these stars.
A distinct subsample ($\sim 7 \%$) of massive OB stars shows evidence for large scale magnetic fields at their surfaces (e.g., \citealt{2013MNRAS.429..398P,2016MNRAS.456....2W}). These fields have been characterised: they are stable on long time scales, and the measured variations in the line-of-sight field strength are well understood under the oblique rotator model. 
In this article we elaborate on a developing consensus how surface magnetic fields can be accounted for in state-of-the-art stellar evolution models. 
\section{Methods}

\subsection{Analytical prescription of surface magnetic fields}

Our understanding of the interactions of the winds of hot star with surface magnetic fields has have greatly benefited from multidimensional magnetohydrodynamic (MHD) simulations performed by \cite{2002ApJ...576..413U} and subsequently by \cite{2009MNRAS.392.1022U}. These simulations have successfully established - in accord with observational evidence - that surface magnetic fields have two major `surface effects'.

The magnetic field lines channel and confine the wind plasma, an interaction that is fundamentally described by the ratio of magnetic energy density and wind kinetic energy density, that is the \textit{equatorial magnetic confinement parameter}, 
\begin{equation}
\eta_\star = \frac{B_{p}^2 R_{\star}^2}{4 \dot{M}_{B=0}v_{\infty}} \, , 
\end{equation}
introduced by \cite{2002ApJ...576..413U}. Here $B_p$ is the polar magnetic field strength, $R_\star$ is the stellar radius, $\dot{M}_{B=0}$ is the mass-loss rate the star would have in absence of a magnetic field, and $v_{\infty}$ is the terminal wind velocity.
\emph{Mass-loss quenching} refers to the phenomenon by which wind plasma is trapped inside the magnetosphere by this channeling, hence the effective mass-loss rate is less than it would be in absence of a magnetosphere. This can be described by a scaling factor, $f_B$, that is, 
\begin{equation}
f_B = 1 - \sqrt{1 - \frac{R_\star}{R_c}} \, , 
\end{equation}
where $R_c$ is the closure radius. As a consequence, the effective mass-loss rate will be
\begin{equation}
\dot{M}_{\rm effective} = f_B \, \cdot \, \dot{M}_{B=0} \, .
\end{equation}
\emph{Magnetic braking} accounts for surface angular momentum removal due to Maxwell stresses and the magnetic field's capability to transport energy and momentum. The additional angular momentum removed on a dynamical timescale is calculated as,
\begin{equation}
\frac{\mathrm{d}J}{\mathrm{d}t} = \frac{2}{3} \Omega R_{\star}^2 R_{A}^2 \dot{M}_{B=0} \, ,
\end{equation}
where $\Omega$ is the surface angular velocity and $R_A$ is the Alfv'en radius. 
The magnetic field strength scales with $~R_{\star}^{-2}$, assuming magnetic flux is conserved at the stellar surface during the evolution of the star.

\subsection{Stellar evolution codes}

Currently there are two hydrodynamic stellar evolution codes that have incorporated the analytical expressions above in order to account for the effects and the evolution of a surface magnetic field. 

Modules for Experiments in Stellar Astrophysics (MESA, \citealt{2013ApJS..208....4P}) is a versatile, open-source stellar evolution code. MESA is very rapidly developing and the code capabilities have greatly increased over recent years. 
MESA models including surface magnetic fields were recently described by \cite{2017MNRAS.466.1052P} and \cite{2017IAUS..329..250K}.
The Geneva stellar evolution code (GENEC, \citealt{2008Ap&SS.316...43E}) has been actively used for over three decades. 
GENEC models incorporate magnetic braking \citep{2011A&A...525L..11M}, as well as the quenching and field evolution components \citep{2017A&A...599L...5G}.

%
%

\section{Results \& Discussion}

A key result from previous studies has been that mass-loss quenching can alter massive star evolution by retaining a significant fraction of the star's mass. As a consequence, magnetic progenitors can explain the existence of `heavy' stellar mass black holes \citep{2017MNRAS.466.1052P} and pair instability supernova \citep{2017A&A...599L...5G} even at solar metallicity. 

\begin{figure}
\centerline{\includegraphics[width=7.5cm,clip=]{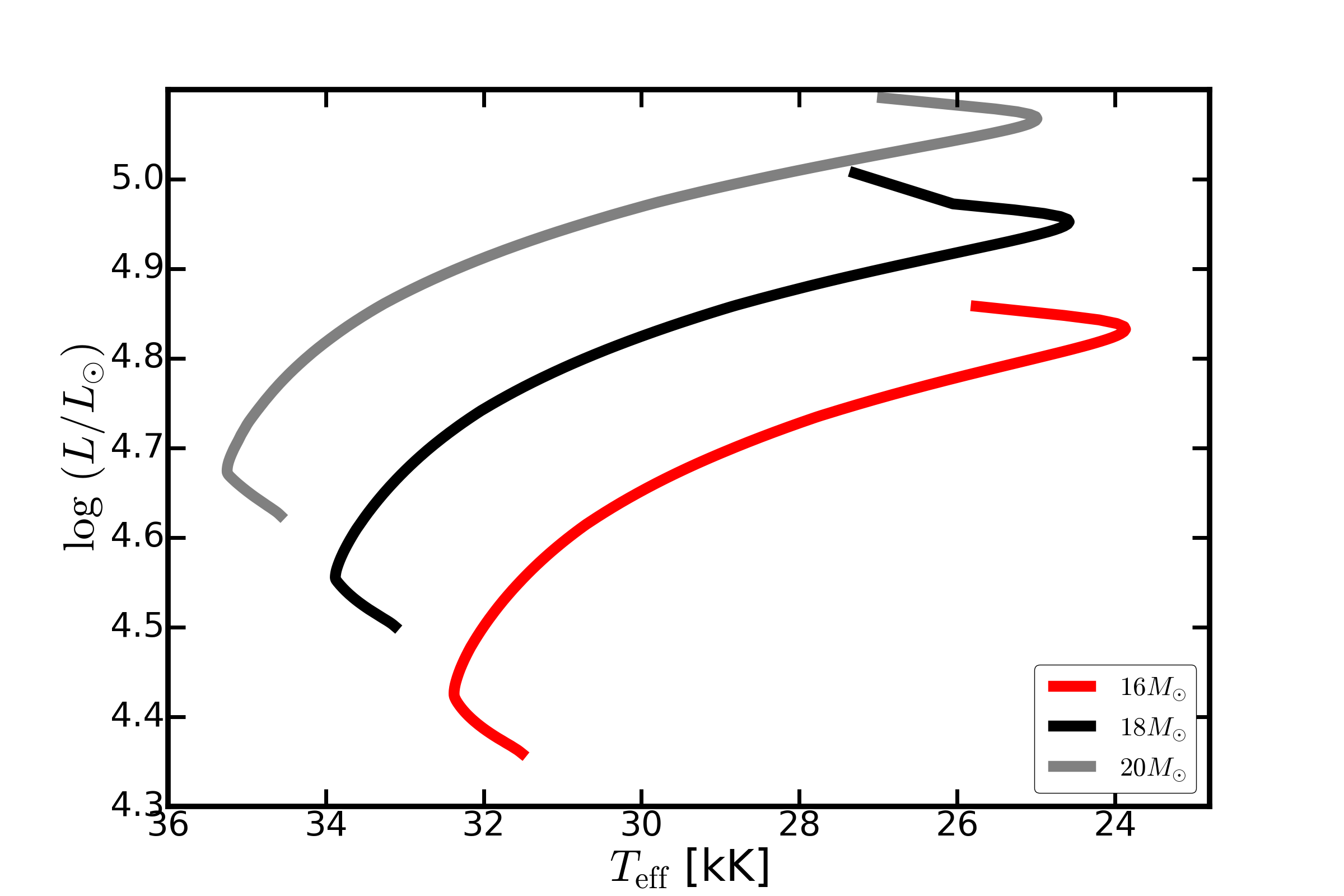}\includegraphics[width=7.5cm,clip=]{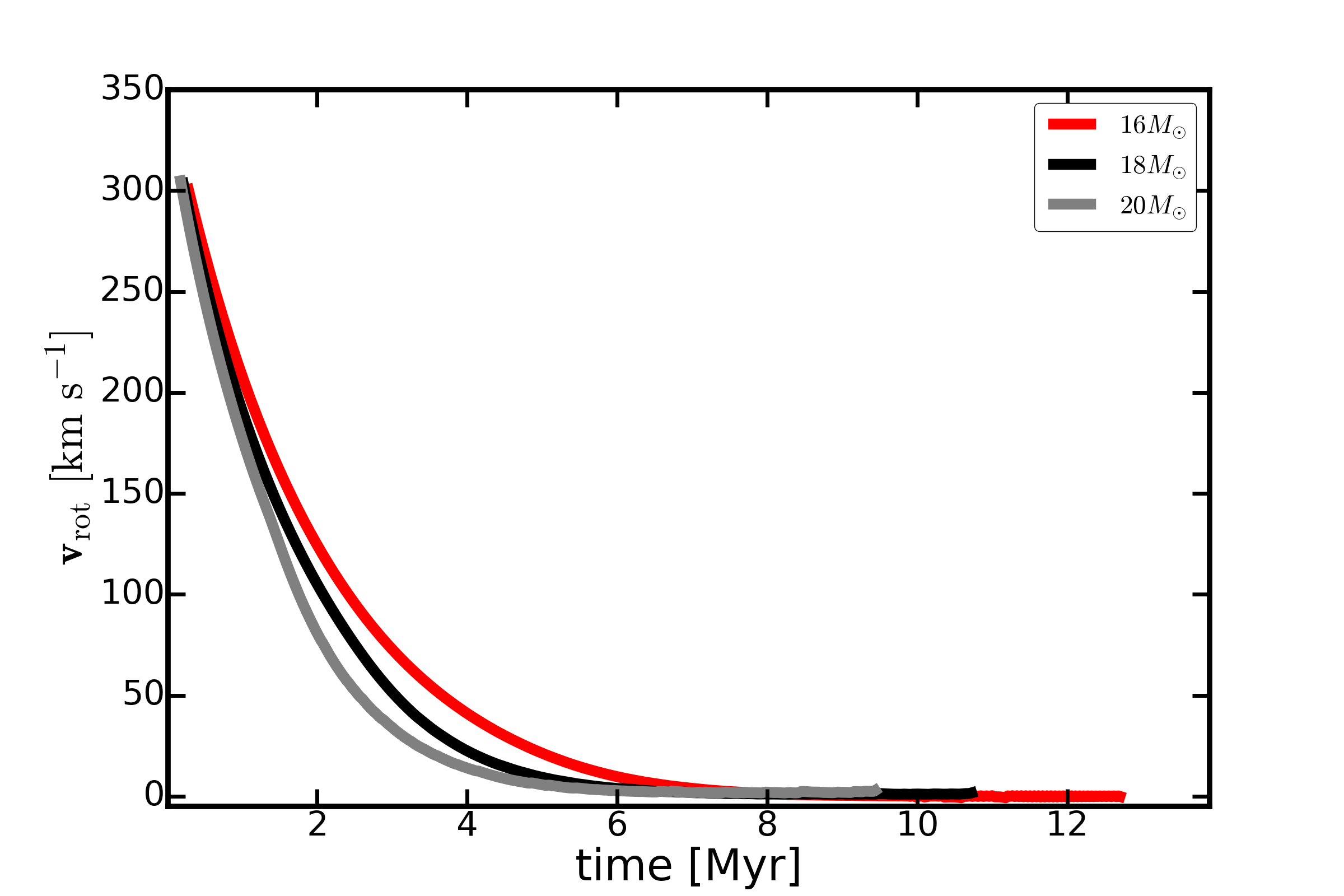}}
\caption{MESA models with $v_{\rm rot} (initial) = 300 \, \mathrm{km \, s^{-1}}$, assuming solid body rotation. These models include the effects of mass-loss quenching, magnetic braking, and magnetic field evolution. \textit{Left panel:} The HRD shows that an initial blueward evolution on the main sequence is followed by a redward turn when the rotational velocity approaches zero. \textit{Right panel:} The time evolution of the surface rotational velocity shows that magnetic braking may take several Myr, and it depends on the initial mass of the star.}
\label{fig:first}
\end{figure}

However, it has became evident that the surface spin-down due to magnetic fields is very sensitive to the model details. To this extent, we computed rotating models at solar metallicity ($Z = 0.014$) with the MESA code. These models include a strong internal coupling between the core and the envelope, which leads to solid body rotation in every model.
In Figure \ref{fig:first} we show main sequence models with initial masses of 16, 18, and 20 $M_{\odot}$ including the effects of mass-loss quenching, magnetic braking, and magnetic field evolution. These models do require several Myrs to completely brake their surface rotation. This depends on the strength of their stellar winds, which is tightly tied to their initial mass.

\begin{figure}
\centerline{\includegraphics[width=7.5cm,clip=]{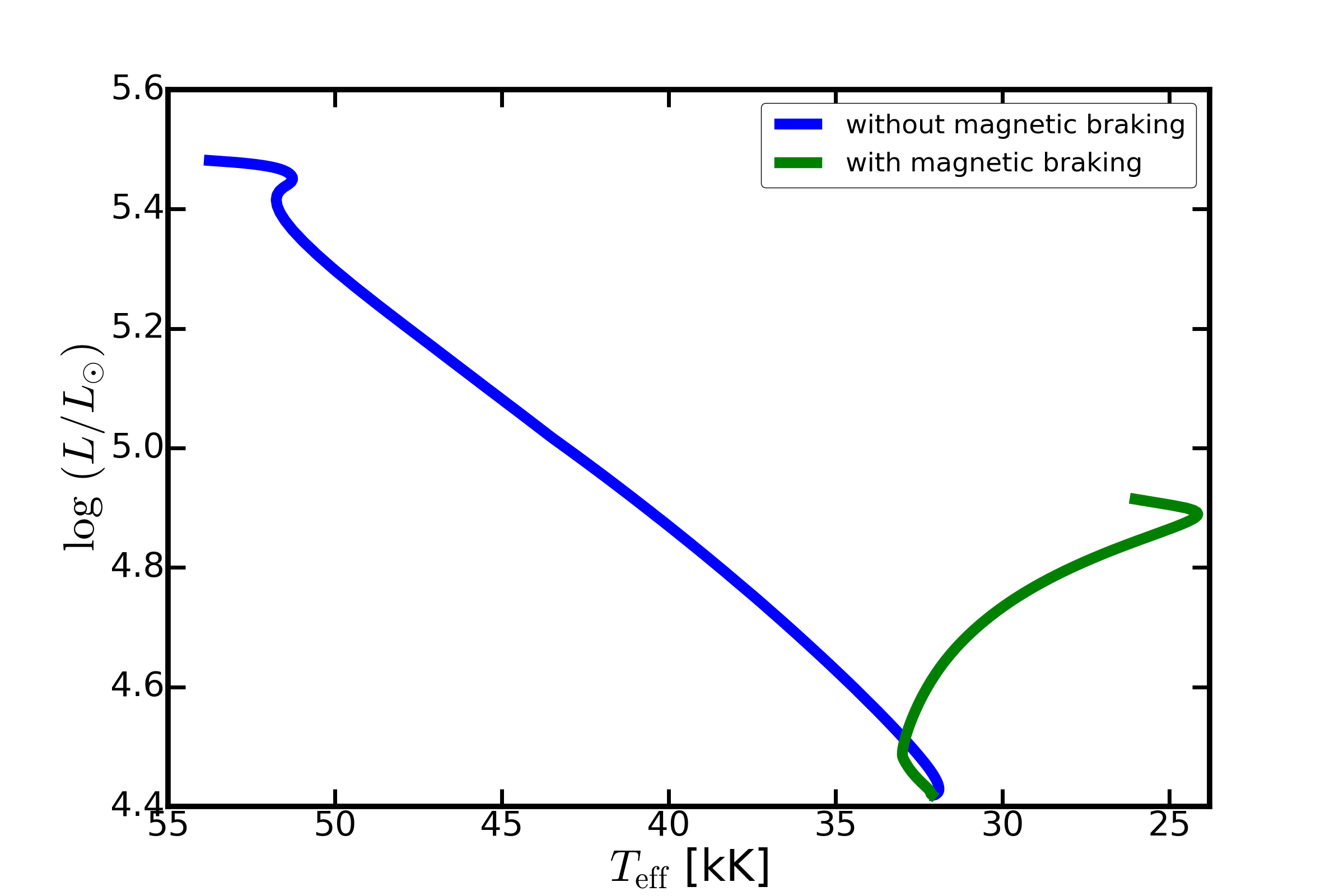}\includegraphics[width=7.5cm,clip=]{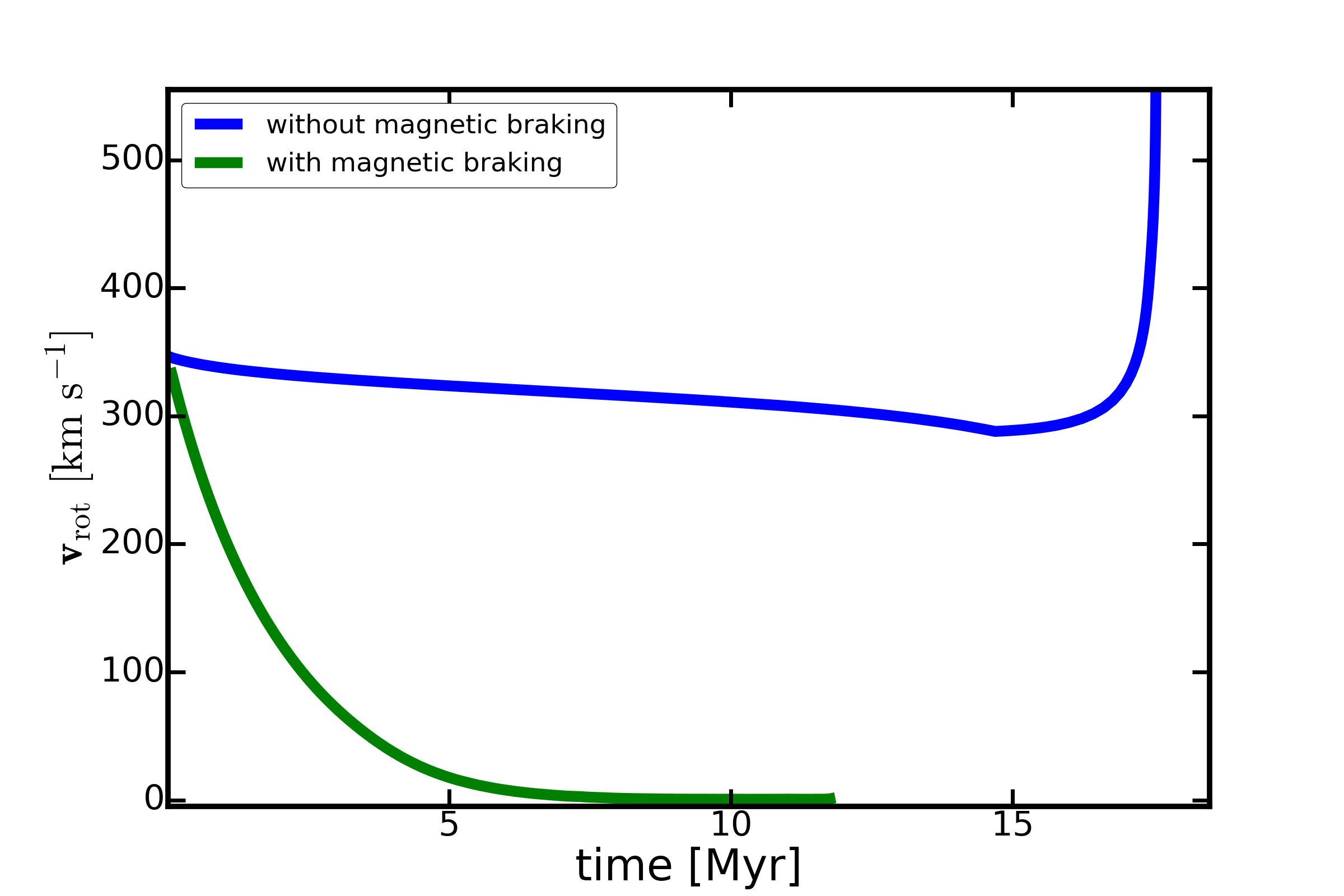}}
\caption{Two 17 $M_{\odot}$ MESA models with solar metallicity. Both models include mass-loss quenching and magnetic field evolution, but one model (green line) includes magnetic braking, the other (blue line) neglects magnetic braking. \textit{Left panel:} The HRD shows large differences between the two models, which originate from their rotational properties. \textit{Right panel:} The surface rotational velocity vs time plot shows that with solid body rotation, the model maintains a constant rotational velocity on the main sequence if magnetic braking is not accounted for.}
\label{fig:second}
\end{figure}

In contrast, Figure \ref{fig:second} shows two models with the same input parameters but the only difference is that one model (green line) is computed with the inclusion of magnetic braking and the other (blue line) is computed without magnetic braking. Mass-loss quenching and field evolution is included in both models. The rotating model with mass-loss quenching but without magnetic braking evolves `blueward', that is, chemically homogeneously. It is indeed expected that models including core-envelope coupling are very efficiently mixed by meridional currents. In the other model, the inclusion of magnetic braking does brake the whole star (it is nearly solid body rotating). This rapidly decreases the meridional currents and thus the mixing of the elements is less efficient. The star evolves `redwards' once the rotational velocity approaches zero. Models computed with a core-envelope coupling and an initial rotation of 300 km s$^{-1}$ evolve homogeneously. With the same conditions, the models evolve redward when magnetic braking is accounted for. This result would mean that in case of a coupled core-envelope configuration, a star evolving redwards either should begin with an initial rotation smaller than the one considered here, or its surface rotation should be braked by magnetic braking. These two scenarios have different predictions for the changes of the surface abundances.

%


\section{Conclusions}

The incorporation of surface magnetic fields in stellar evolution models has resulted in identifying two new evolutionary pathways of massive stars. Additional channels may also be discovered, however, this will require new and extensive grids of models extending the parameter space of previous studies. Currently, the model dependence of the inclusion of magnetic braking needs to be thoroughly investigated.

Forthcoming works will focus on how state-of-the-art stellar evolution models will allow the improved derivation of stellar parameters of observed magnetic stars, how the observables (e.g., surface nitrogen abundance, rotational velocity) evolve in these models, and for how long the magnetic confinement can be maintained during the evolution.

\acknowledgements
We appreciate fruitful discussions with Rich Townsend and Alexandre David-Uraz.
GAW acknowledges support in the form of a Discovery Grant from the Natural Science and Engineering Research Council (NSERC) of Canada.
VP acknowledges support provided by the NASA through Chandra Award Number GO3-14017A issued by the Chandra X-ray Observatory Center, which is operated by the Smithsonian Astrophysical Observatory for and on behalf of the NASA under contract NAS8-03060.
\bibliography{demo_caosp305}
\end{document}